# A First Look at Firefox OS Security

Daniel DeFreez*, Bhargava Shastry[†], Hao Chen*, Jean-Pierre Seifert[†]
*University of California, Davis
{dcdefreez, chen}@ucdavis.edu
[†]Security in Telecommunications, Technische Universität Berlin
{bshastry, jpseifert}@sec.t-labs.tu-berlin.de

*Abstract*—With Firefox OS, Mozilla is making a serious push for an HTML5-based mobile platform. In order to assuage security concerns over providing hardware access to web applications, Mozilla has introduced a number of mechanisms that make the security landscape of Firefox OS distinct from both the desktop web and other mobile operating systems. From an application security perspective, the two most significant of these mechanisms are the the introduction of a default Content Security Policy and code review in the market. This paper describes how lightweight static analysis can augment these mechanisms to find vulnerabilities which have otherwise been missed. We provide examples of privileged applications in the market that contain vulnerabilities that can be automatically detected.

In addition to these findings, we show some of the challenges that occur when desktop software is repurposed for a mobile operating system. In particular, we argue that the caching of certificate overrides across applications—a known problem in Firefox OS—generates a counter-intuitive user experience that detracts from the security of the system.

## I. INTRODUCTION

In July 2013, Mozilla launched Firefox OS, an entirely web-based mobile platform. Primarily aimed at developing markets, Firefox OS devices tend to be low-powered, but inexpensive, and therefore offer an attractive alternative for consumers that cannot afford an Android or iOS device [29]. The lure for developers is, of course, the possibility of creating applications that conveniently work across platforms. One of the primary barriers to the adoption of web-based applications has always been lack of hardware access. Mozilla is attempting to solve this problem by leading a standardization effort for new Web APIs [7]. Meanwhile, Firefox OS provides a glimpse at what the future may bring for the mobile web.

Being a late entrant in the smartphone market, Mozilla has had the advantage of hindsight. Firefox OS adopts many of the security features of other mobile platforms, while avoiding some of the missteps. Laying at the intersection of the web and mobile security models, Firefox OS maintains the same-origin policy while adopting the use of applications as security principals that has become the hallmark of mobile systems. Not only are Firefox OS applications isolated from one another, but even remote code from a single origin that spans applications is sandboxed, as if the applications were running in separate browsers. This model prevents some of the issues that have plagued Android [37], and serves as a partial realization of previous recommendations regarding the interplay between web and mobile security [9], [31].

While the platform itself is relatively robust, application developers make plenty of mistakes. Here, too, Mozilla has learned from its forebears, taking steps to proactively contain the abuse of vulnerable applications. All privileged applications have a default Content Security Policy applied and are reviewed for conformance to security guidelines prior to being distributed through the official Firefox Market. The Content Security Policy almost categorically prevents Cross-Site Scripting (XSS) via JavaScript injection, and code review should pick up any misuse of permissions or obvious security errors. This paper asks whether these mechanisms are sufficient to prevent developers from making trivially preventable security blunders. We find that they are not. The most prevalent attack vector, without a doubt, is HTML injection, and `.innerHTML`[1] is the culprit. The `.innerHTML` property parses a text string and replaces a node's content with the HTML representation of that string. It is enormously popular due to its convenience, and is often called behind the scenes in libraries such as jQuery. Mozilla has taken great strides to eliminate the use of `.innerHTML` and friends in its official apps, but it is still allowed in the market. We also find several other classes of security errors made by applications in the market. Moreover, in section III we show that these vulnerable applications are automatically detectable with lightweight static analysis.

Adapting desktop browser technologies to a mobile security environment is bound to introduce complications. Our observation is that the act of sandboxing applications creates an expectation that the side effects of any action will be restricted by application boundaries. As such, any behavior that deviates from this expectation represents a security risk. One example of such dissonant behavior is the way Firefox OS handles certificate overrides. It is a known issue that in Firefox OS certificate overrides for a domain are applied across applications, rather than on a per-application basis. When combined with the limited UI available in Firefox OS surrounding connection security, system-wide caching of certificate overrides puts users at risk and presents a counter-intuitive user experience.

*Contributions*: We provide the following contributions:

1) We find that developer practices in the market frequently violate Mozilla's security guidelines for applications, including the direct insertion of user input into the DOM, failure to properly handle the origin of web messages, and the use of HTTP instead of HTTPS.
2) We demonstrate that lightweight static analysis can be used to help find vulnerable applications in the Firefox Marketplace.

---

[1]There are several other properties that are also problematic, such as `.outerHTML`.

3) In light of the certificate override caching problem, we look at the consequences of retrofitting web applications on legacy system software. We document discrepancies in Firefox OS' user interface and demonstrate how the caching problem and inadequate security indicators pose a security risk.

## II. Background

Firefox OS is divided into three layers: Gonk, Gecko, and Gaia. Gonk is the Android-derived underlying OS, Gecko is the layout engine (including SpiderMonkey, which provides JavaScript, and several other legacy sub-systems), and Gaia is the UI. All applications are written with common web technologies (HTML5, CSS, JavaScript) and run on top of Gecko. Even Gaia, which provides the core set of applications shipped with a device, is built as a set of HTML5 apps. Gecko enforces permission-based access control over applications' access to device APIs.

Firefox OS applications are divided into three security categories: unprivileged, privileged, and certified:

- Unprivileged applications have a restricted set of permissions available. They can be either hosted or packaged. Hosted applications are "installed," but the content for the application is hosted elsewhere. A packaged application has all of its content placed placed in a zip file which is distributed through a market, and the market is responsible for signing the package. A manifest is included in the package which contains meta-data about the application, including permission requests.
- Privileged applications have access to a richer set of possible permissions. They also have access to additional features such as the ability to declare local redirects. Because of the increased level of trust, privileged applications are required to be packaged, have a default Content Security Policy applied, and must go through a code review process prior to being accepted into the market.
- Certified applications are installed as part of Gaia and have access to all permissions.

Any application can embed web content. In Firefox OS, where every piece of the user interface is within a "browser" of sorts, the distinction between the app and web content is that web content is viewed within an `iframe` and comes from a different origin. The security restrictions for web content are very similar to web content in a desktop browser, and the permission set is the most restrictive of the categories.

While hosted and unprivileged packaged applications have the same set of permissions available, there is an important difference in the way the two types of apps are installed. The `.zip` archives for packaged applications are housed by the market. They are signed by the market and delivered over HTTPS, thereby ensuring the integrity of the application. Naturally, packaged applications can only be updated through the market, allowing the user a degree of control and potentially the ability to audit the code of the app. Hosted applications, on the other hand, behave more like bookmarks. Only the application manifest is fetched at install time, which we find is almost always delivered over HTTP rather than HTTPS.

Running a hosted application implies fetching the application content from a remote server, preventing any user control over the content of the application and allowing permissions to be tampered with at the time of installation. Fortunately most of the sensitive permissions available to hosted applications prompt upon first use.

As outlined in the official documentation [3], Firefox OS uses a small set of permissions. There are 56 permissions in the permissions table, of which only 23 are usable by uncertified applications. Table I provides a list of all permissions requested in the market. The most popular permission, `systemXHR`, enables the use of `XMLHttpRequest` across origins.

In contrast to Android, Firefox OS permissions can be either implicit or explicit. Implicit permissions are granted at install time if the application requests the permission in its manifest. Explicit permissions trigger a prompt at the time of first use, and are revocable through the settings app [26]. The security category that an application belongs to determines which permissions are implicit and which are explicit. For web content, all available permissions are necessarily explicit, which at the time of writing only includes `desktop-notification` and `geolocation`.

Just because a permission is requested does not mean that an application actually uses that permission. There are 35 applications, for example, that request the systemXHR permission but are incapable of performing cross-origin XHR. This is identified by looking for `mozSystem: true` as a parameter to `XMLHttpRequest`. In total, we find a lower bound of 50 applications in the market that request permissions for which they do not use the corresponding web API. Use is determined by a permission map that was generated from the Firefox OS source code by looking for calls to the permission manager. This may be an underestimate, because it is assumed that all code is reachable. Having overpermissioned apps is undesirable from the standpoint of the principle of least privilege, and demonstrates the imprecision of manual review, but it is unclear how an application with extra permissions could be coerced into using its permissions without being able to inject JavaScript.

| Permission | Apps | |
|---|---|---|
| systemXHR | 142 | (24.9%) |
| geolocation | 106 | (18.6%) |
| storage | 54 | (9.5%) |
| desktop-notification | 53 | (9.3%) |
| device-storage:sdcard | 51 | (8.9%) |
| browser | 31 | (5.4%) |
| audio-channel-content | 28 | (4.9%) |
| device-storage:pictures | 27 | (4.7%) |
| alarms | 21 | (3.7%) |
| contacts | 16 | (2.8%) |
| tcp-socket | 10 | (1.8%) |
| mobilenetwork | 7 | (1.2%) |
| device-storage:videos | 6 | (1.1%) |
| 8 other permissions | 17 | (3.3%) |
| Total | 570 | (100%) |

TABLE I. Permissioned Apps in the Firefox OS Market (February 2014)

At the time of writing, we are not aware of any permissions abuse or other malware in the Firefox OS market. While this is undoubtedly a result of Firefox OS' nascent state, the use of code review for privileged applications is also a deterrent. Malware will almost certainly surface in the future. In anticipation of this, Mozilla has attempted to minimize the damage that a malicious application could do. Each application runs in a separate Gecko child: a *Content* process which provides a sandboxed execution environment. This includes separate cookies and storage, regardless of origin. The possibility of escape from this sandbox is limited because no facility is provided to run native code. Instead, hardware access is provided through JavaScript APIs. Should an application find a way out of this sandbox, Firefox OS will soon utilize *seccomp* to filter the system calls available to application processes. In addition to the threat of privilege escalation, the availability on Android of native operating system facilities - such as the ability to list running processes - has led to information leakage through side channels [39]. Firefox OS, in contrast, provides very limited access and protects much of this information with certified permissions.

From an application security perspective, the most important defense is the the introduction of a default Content Security Policy (CSP) for privileged applications. Content Security Policy is a mechanism for defining the allowed origins of resources [34]. The most important feature of the default policy is that it disables third-party and inline JavaScript. Disabling inline JavaScript means that not only will inline `<script>` tags be ignored, but also `eval`, some incantations of `setInterval`, and other functions that interpret strings as code. Fortunately, the default CSP was introduced early on in the development of the Firefox OS ecosystem. Bringing already developed applications into compliance can require some effort, and we suspect this will be an issue for a number of developers as the number of privileged APIs available increases. CSP does nothing, however, to prevent the broader problem of HTML injection. As [18], [38] discuss, and as we shall see in the next section, this still leaves an attacker room to maneuver.

*Advertising:* It has been observed [35] that the lack of privilege separation between Android applications and advertising libraries can lead to permission abuse. Applications and advertising libraries exist in a state of mutual distrust, and a number of papers have looked at the possible ways of separating advertising libraries from applications [28], [30]. In contrast to Android, advertising on Firefox OS closely follows the web model, which has the potential for better isolation between applications and advertisements. Web advertisements are often isolated by placing them in an `iframe`. This isolation can be made more robust through the use of the `sandbox` attribute on the frame, guaranteeing a unique origin (and thus a different set of permissions), as well as preventing forced navigation attacks. Unfortunately, few Firefox OS applications go through the effort of sandboxing advertisements. The reference implementation from Mozilla [36] places the integration code for the advertisement in an unsandboxed `iframe` that is sourced from a `data-URI`. The advertisement and several remote scripts - including a fingerprinting script - are thus loaded in the same origin as the app. Any permissions that an application[2] declares are fair game for the ad provider to use. So if the user has granted, say, `audio-capture` to the app, it has also been granted to the ad provider. This is actually a higher risk situation than on Android: the code changes without warning, and an active network attacker could abuse any permissions granted to the application.

---

[2]This integration library is targeted at hosted, rather than privileged apps, but this still leaves a number of permissions available.

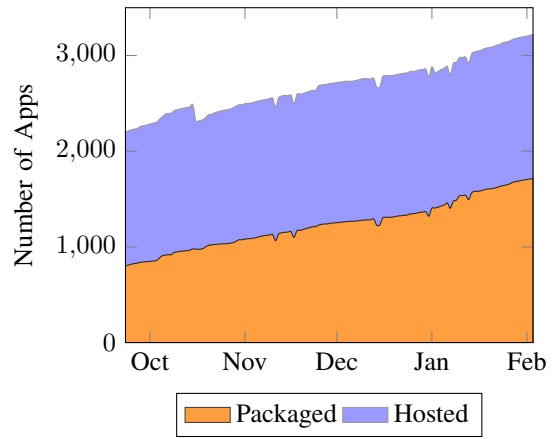

Fig. 1. Firefox OS Market Size, 2013-2014

## III. DETECTING VULNERABILITIES IN FIREFOX OS APPS

This section identifies potentially vulnerable apps through lightweight static analysis. The vulnerabilities reported here are taken from applications in the Firefox Market. One challenge is that the Firefox market is so much smaller than its counterparts. While Google Play and the iTunes App Store each have over a million applications, the Firefox Market has a mere 3,000. Figure 1 shows the size of the Firefox OS market between October 2013 and February 2014, as measured by the number of application manifests in the official tarballs made publicly available by Mozilla. Surprisingly, the growth of the market is driven almost entirely by packaged applications, which as of January 2014 outnumber hosted applications in the market.

We are primarily concerned with analyzing privileged apps, which further constricts the dataset. From a code review perspective, hosted applications are not significantly different than web applications. There is little incentive to review applications that can be continuously updated. This leaves us with 570 applications. The fact that we are able to find multiple vulnerable applications by using relatively simple techniques suggests that the Firefox Marketplace is a promising area for further vulnerability research.

Precise static analysis of JavaScript is a notoriously difficult problem [19] due to the dynamic nature of the language, and most tools have difficulty scaling to large programs (though progress is being made [33]). Our approach eschews precision in favor of efficiency. While our analysis is neither sound nor complete, it establishes that heuristics can be applied and still find vulnerabilities in the market. Whereas other techniques might timeout analyzing a single program (see Section VI), our tool is capable of scanning the entire market in a matter of minutes on a standard desktop machine. Using just a few heuristics we are able to flag several applications as vulnerable. Two of these are privileged, and therefore have already undergone code review. These applications are presented in Section IV.

Mozilla already has an open source tool for validating applications that are submitted to the market, called simply "the validator." The validator is used to help code reviewers analyze privileged applications before they are accepted into the market, and Mozilla has provided a web interface where developers can run it against their own applications. The validator primarily identifies errors that will break an application once it is installed, such as manifest inconsistencies or the use of JavaScript functions that violate the CSP. The validator does warn when properties such as `innerHTML` are assigned non-literal values, but this is such a frequent occurrence, and the rate of false positives for injection attacks is so high, that it is difficult to see how a reviewer could do anything but ignore these warnings.

The market validator would be natural location for additional security checks, and an early version of our tool was simply a set of extensions to the market validator. Based on feedback from Mozilla, however, it was determined that a central static analysis tool was too computationally expensive to be practically useful. As such, our current tool stands separate from the validator and is built on top of the Acorn [22] JavaScript parser and the Tern [23] type inference engine. The choice of technologies was influenced by Mozilla's ScanJS project (see section VI), so as to facilitate the possibility of incorporation into that project. Our tool, however, shares no code with ScanJS. At the time of writing, ScanJS is not capable of finding the vulnerabilities discussed here.

The techniques described in the sections that follow build upon a common global analysis that is performed for each application. First, def-use chains are built for every variable. These are created through a limited, context-sensitive inter-procedural analysis. It is limited in the sense that it does not yet handle a number of aspects of JavaScript, such as higher-order functions, and thus makes several approximations.

After the def-use chains have been created, uses for known sinks, sources, and filters are resolved to abstract definitions such as "sink" and "filter," and the use entries are updated to reflect this. This list of sinks, sources, and filters includes the standard set of properties available in most XSS references, and also a set of popular library functions that are known to achieve the same effect, such as the `.html` function in jQuery.

For every variable that is used at an interesting location such as a sink or filter, we use Tern to infer the type of that variable. Tern is normally used to provide auto-completion in an editor, but is also well-suited to our purpose. We require fast, but not necessarily perfect type resolution. JavaScript is a classless language, and as such determining when two objects are of the same "type" is difficult. Tern makes headway by identifying common programming patterns. One example given in the Tern documentation [23] is that of type extension. Type extension in JavaScript is often achieved by writing a utility function that copies an object prototype and then adds properties to this new object. Understanding the relationship between objects created in this manner necessarily requires guesswork, and Tern provides a set of heuristics for making educated guesses in these situations.

Next, constant propagation is performed when a variable is used as a property name. This allows us to produce reports in spite of certain obfuscation techniques. One such case in the market is the popular ConnectA2 app, which hex-encodes all

```
window.addEventListener('message', function(evt) {
  authWindow.close();
  token = evt.data.token;
  self.exchangeToken();
});
```

Fig. 2. Example message handler that fails to validate the message origin

```
window.addEventListener('message', function(evt) {
  if (evt.origin !== 'app://example.com') {
    return;
  }
  var data = evt.data;
  token = evt.data.token;
  self.exchangeToken();
});
```

Fig. 3. Correct validation of message origin

object properties. Then, after the def-use chains have been built and the types of variables have been inferred, a set of policy rules are applied. These rules are described in the sections that follow.

### A. Origin Validation

The approach used here is to conservatively flag uses of the messaging API that do not validate the origin of messages by identifying instances where the origin[3] property is not accessed at all. If an application is going to validate the origin of a message, as in Figure 3, the origin property must be read somewhere. The exact conditions used to identify apps that fail to validate the origin of web messages are as follows. The app must:

1) Register a message event handler
2) Not read the origin property in any scope where the message event is alive

The first condition is met by traversing the Abstract Syntax Tree (AST) and looking for a call expression to `addEventListener` which passes the string literal `message` as the first parameter.

The second condition is evaluated by analyzing the body of the message handler. The first parameter to the message handler, `evt` in Figure 2, is the actual message. The def-use chains for the `evt` variable are searched for accesses of the origin property. If the origin property for the event is never accessed, then it is impossible for the app to verify that the origin is correct. Note that false negatives are obviously possible. We only identify cases where the applications *could* validate the origin of the message, not where it does.

It is has been shown in [32] that many of the top websites fail to correctly validate the origin of messages. Similar results are found here, despite the fact that the use of `postMessage` is featured prominently in Mozilla's security guidelines. Of the 83 applications that register message handlers, 56 of them fail to validate the origin of messages. We make no claim that these applications are vulnerable to attack. It may be that there is no way for external content to send messages to these apps. Section IV, however, demonstrates that under the right conditions, failure to validate the origin of a message event can have dire consequences.

---
[3]We also look for the source property, but here we refer to the two collectively as the "origin" property.

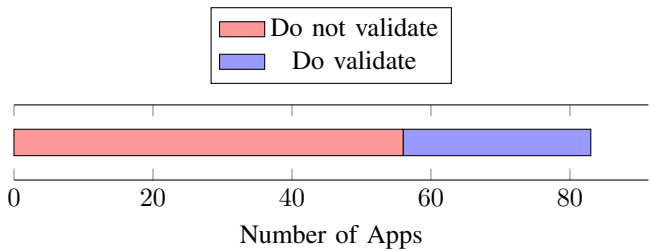

Fig. 4. Message origin validation in Firefox OS Market

## B. HTML Injection

The most straightforward way to discover applications that are possibly vulnerable to HTML injection is to list all applications that pass a variable, rather than a constant, to a sink capable of inserting raw HTML into the Document Object Model (DOM). This is the criteria used by the market validator. The problem with this approach is that most of these uses are not exploitable. Banning the use of these dangerous properties entirely would be one solution, but it seems unlikely that Mozilla is going to take that step.

The approach used here is to detect instances where the same type of data is used more than once in an application, but not filtered in all uses. The observation is that developers filter data when they believe an attacker can control the input. If they also use that type of data without filtering it, then this is likely a mistake. Specifically, an app will be flagged if:

1) Data from a defined source is read into reaching definition `v1` of inferred type A
2) `v1` is passed to a defined sink
3) Data of inferred type A is filtered, creating reaching definition `v2`
4) `v2` is passed to a defined sink

Note that the only relationship between `v1` and `v2` is that they have the same inferred type. This is a potential source of false positives, as it inherits the imprecision of the type inference process.

The first condition requires the development of a list of sensitive sources. This list was developed by hand, and includes XHR, traditional DOM XSS sources, filenames, web messages, etc. The second and fourth conditions are established in similar fashion. A standard list of sensitive sinks documented by Mozilla is used, along with a custom list of library-specific sinks such as the `html` function in jQuery. The third condition is met by a heuristic. If the `replace` method is used, or if a temporary DOM element is created and the `textContent` of that element is immediately read, the data is considered to be filtered. This is meant to be merely a first step toward identifying the way applications filter data.

Consider the example in Figure 5. The tool makes the assumption that when JavaScript functions are used in an object-oriented fashion, i.e. functions are called with the constructor invocation pattern, then all instances of an object with that prototype can be treated as if they were the same object. They are not actually the same object, of course. Therefore it is possible that one instance may need filtering while another may not, but this is rare in practice. In the example, all instances of `Something` are treated as if they are the same variable and

```
function unfiltered(v) {
    var html = '<div>' + v.getName() + '</div>';
    element.innerHTML = html;
}

function filtered(v) {
    var filtered = filter(v.getName());
    var html = '<div>' + filtered + '</div>';
    element.innerHTML = html;
}

function filter(s) {
    return (s || "").replace(/</g, '<');
}

var data1 = new Something();
var data2 = new Something();
unfiltered(data1);
filtered(data2);
```

Fig. 5. Inconsistent filtering of attacker controlled input

tracked accordingly. In the `unfiltered` function, a property of this single variable is assigned directly to `innerHTML`, a sensitive sink. The tool identifies the fact that the same property is used in the `filtered` function. It is first passed as a parameter to `filter`, where `replace` is called. Then it is used in a sensitive sink.

Five applications vulnerable to HTML injection were detected using the method of looking for filtering inconsistencies. Manual analysis showed that only two of these were legitimate injection vulnerabilities. One application intersected the HTML injection and origin validation sets, as discussed in section IV. Two other applications have been identified as vulnerable to HTML injection by manual analysis.

## C. Use of plaintext HTTP

We find that it is common for privileged applications to use HTTP instead of HTTPS for XHR communication. The use of HTTP instead of HTTPS is identified by constant propagation. Whenever the `open` method of an `XMLHttpRequest` object is called, our tool looks at the second parameter. The app is flagged with a low-priority warning if this is a string literal that starts with `http://`. The ConnectA2 application, which is one of the most popular apps in the market and consistently on the front page, was flagged in this fashion, despite obfuscation. During registration, ConnectA2 sends the phone number of the registrant in the clear to their servers. While not a serious privacy violation, this is certainly undesirable. Our results show that 48 apps open systemXHR connections over insecure connections, which is just under 50% of the applications that actually use systemXHR.

## IV. CASE STUDIES

This section presents three vulnerable applications discovered in the Firefox OS market. The first is an unprivileged application susceptible to a traditional XSS attack. This demonstrates that unprivileged Firefox OS applications are on the same security footing as normal web applications. The second app is privileged, but susceptible to HTML injection via local filenames on the device. The third app is privileged and susceptible to oAuth session forgery despite CSP due to a failure to validate the origin of web messages combined with HTML injection.

*Classic XSS:* This is an exploitable vulnerability that was manually discovered in a version control application prior to static analysis. Our tool is capable of finding the vulnerability automatically. This app allows a user to access their Github account to view Gists. The application has an XSS vulnerability caused by failing to sanitize the names of Gists. This allows an attacker to inject JavaScript. Without CSP it is trivial to extract login credentials and exfiltrate them to an attacker controlled site. Only privileged apps are protected by a Content Security Policy. This leaves a large swath of the packaged market potentially susceptible to classic XSS attacks.

*HTML Injection via Filenames:* This is a vulnerability that was manually discovered in a privileged document editing app prior to static analysis. Our tool is capable of finding the vulnerability automatically. Many apps do not expect that HTML entities can be put in filenames, but most HTML characters other than the forward slash are valid filename characters in Linux. The document editing app reads from the SD card on the device. A malicious app on the same device could create filenames containing HTML entities. When the the target application reads the list of files it is susceptible to HTML injection because the filename is used directly with `innerHTML`. The possibility of exploitation is limited, however, as the malicious app cannot close tags and the target app has no other privileges.

It is unlikely that any application will rely on filenames such as ``. Since all filesystem access is mediated through the storage web API, which already puts restrictions on filenames, it is recommended that the storage API perform rudimentary HTML entity filtering.

*oAuth Session Forgery:* This vulnerable application was flagged automatically, without prior knowledge of the vulnerability. It is the sole application in the dataset that was identified as an overlap of HTML injection identified through inconsistent filtering and failure to validate the origin of web messages.

One of the most popular applications in the Firefox OS market is a note taking application that syncs with a cloud service. This application uses oAuth 1.0 for authorization. After authorizing the app, the user is redirected back to the application with the request token as a URL fragment via a special redirect declared in the application manifest.[4] This redirect page uses `postMessage(result, *)` to push the request token back to the main application. The use of the wildcard origin here is dangerous. An attacker could receive the now authorized temporary credentials, but this would be difficult to exploit because of the race to exchange this token for a full access token. The message handler which receives the request token is equivalent to Figure 2, i.e. it does not validate the origin of the message it receives.

The application is also vulnerable to HTML injection via the name of the note. This was automatically detected by our tool because the name of the note is used in more than one location, but not filtered before every use. The code for displaying the name of the note is roughly equivalent to Figure 5.

This application has been manually confirmed as exploitable in the following scenario. An attacker crafts a malicious note with an `<iframe>` in the name of the note, with the frame source being a page under the control of the attacker. The attacker then shares this note with the victim. The attacker's page will be opened if the victim accepts the note and views the notebook containing the note with the Firefox OS app. This page can be made invisible and uses `postMessage` to send a valid request token for the cloud service to the app, which the attacker previously obtained by authenticating to the cloud service but not completing the token exchange.[5] The app will exchange the attacker's temporary token for a full access token and replace the access token in its local database with this new token. The next time that the application syncs to the cloud service it will synchronize local notes to the attacker's account.

This example serves as a strong endorsement of the recommendation put forth in [17] to extend CSP to provide messaging whitelists. Indeed, manual analysis of the market shows that most applications need only to send messages to one origin, and in many cases that origin is the *application itself*. The whitelists would be trivial to construct for these applications, and it would prevent an entire class of attacks.

## V. Counter-intuitive User Experience

While the previous section studies multiple classes of security vulnerabilities at the application layer, this section scrutinizes the problem of system-wide caching of certificate overrides. We describe what happens when system services that used to be tightly coupled to the browser program are exposed to applications in a mobile environment without modification. Alongside insufficient security UI in Firefox OS, we demonstrate how the caching problem poses a security risk.

### A. SSL Certificate Caching

The certificate caching problem—a known issue in Firefox OS [11]—is the following: Manually overriding a certificate warning for a web origin not only overrides the warning for subsequent visits to that web origin, but also makes the override applicable to any application on the phone, as though applications queried a shared system-wide cache for certificate overrides. While application code from the same web origin is sandboxed across applications in Firefox OS, the handling of overrides presents an anomaly. We digress to present a simplified flow of SSL certificate validation and override in Gecko. Subsequently, we analyze the problem of certificate caching in detail.

*Certificate Validation and Override Services:* Gecko relies on the Network Security Services (NSS) library [4] for security related services. NSS' SSL/TLS certificate validation service is responsible for validating the SSL certificate chain presented by a remote server in the process of initiating an HTTPS connection. The certificate override service, tied to the certificate warning interstitial page, allows for manual overrides of certificate warnings. It maintains a record of certificate overrides that have taken place in the past, in the form of {Web Origin, Certificate Fingerprint} key-value pairs.

---
[4]These redirects are limited in scope to the application. Only four applications in the market make use of these redirects, and all of them do so for oAuth.

[5]This is complicated by the fact that temporary credentials have a fairly short lifetime. The attacker needs to regenerate request tokens approximately every 30 seconds.

For a temporary override, the key-value pair is cached in Gecko's program memory. When certificate validation for a server fails, the validation service queries the override service to check if an overridden certificate for the server has been cached. If there is a cache hit, the override service verifies if the cached certificate is the same as the certificate presented in the ongoing SSL handshake. If this succeeds, the HTTPS connection proceeds without a warning. Otherwise, the user is presented with a certificate warning interstitial.

A valid certificate vouches for the authenticity of a remote security principal and forms the basis for trust in Internet communications. The browser web and the smartphone platform entertain different notions of security principals. Next, we see how this leads to a confused deputy scenario in Firefox OS.

*Security Principal:* Desktop browsers routinely interact with off-device security principals. The same origin policy establishes a means to identify them. While the policy is reflected in legacy NSS code, the notion of an on-device security principal is something that is alien to security services in the NSS library in general, and the certificate validation and override services in particular.

While Firefox OS relies on the underlying OS kernel for isolating on-device principals, it delegates certificate validation and overriding to the legacy NSS library. So, while the application is treated as a security principal in managing session tokens, local storage, and permission-based access control, the web origin dictates how SSL certificates are handled. This semantic difference between the NSS port in the desktop browser and in Firefox OS makes both the certificate validation and override services, confused deputies. Since the two services still treat web origin as the security principal, certificate validation requests for the same web origin *across* Firefox OS applications elicit the same response.

Next, we examine architectural differences between the desktop browser and Firefox OS that lead to side-effects being distributed and subsequently retained across process boundaries.

*Process Model:* The desktop browser and Firefox OS are built on different process models. The desktop browser program itself and all of its (web) content run in a single OS process. In contrast, Firefox OS is built on a multi-process model. Each web app on Firefox OS runs in a separate *Content* process. Gecko and its constituent sub-systems run in a privileged process called the boot2gecko (b2g) process.

Because the desktop browser and its content pages run in a single process, all transient side-effects are contained within the process' memory; this includes temporary certificate overrides. On Firefox OS web apps running in content processes are untrusted principals, but the b2g process belongs to the Trusted Computing Base (TCB). Security sensitive side-effects such as certificate overrides are registered in the b2g process, even if the operation responsible for the side-effect originated in a content process. The validity of the override stretches through the lifespan of the b2g process. Given that the b2g process is one of of Firefox OS' core components, it is killed only on device restart.

A defining characteristic of the desktop user experience is users being able to start and close (kill) programs (OS processes). This observation combined with the fact that desktop users are exposed to the browser program in its entirety (program = process) instills the notion that side-effects are strictly tied to program lifespan. The smartphone user experience on Firefox OS is starkly different. Since program (process) management of the b2g process is not exposed to smartphone users, user expectations around side-effects are not respected by the underlying OS.

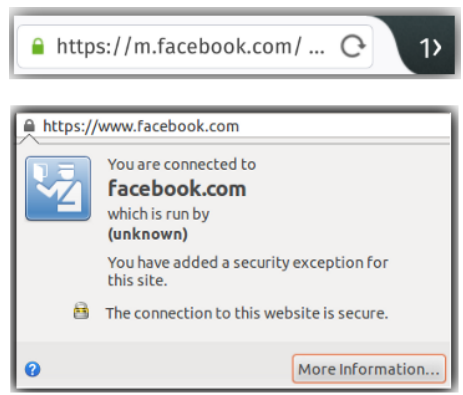

Fig. 6. Site identity button after a certificate override: (Top-Bottom) Firefox OS browser, Firefox desktop browser

Put together, application agnostic certificate handling and retention of overrides in a core process pose a security risk. Next, we see how the problem is exacerbated by discrepancies in the security UI of Firefox OS.

*UI Discrepancy:* Although the caching of temporarily overridden certificates is a characteristic of the desktop browser, there are checks in place to ensure that an active attacker has limited leverage. Temporary certificate overrides are removed when the user closes the browser program. However, a browser restart is not a necessary prerequisite to clear overrides. Desktop Firefox exposes the option[6] to remove overrides during a running instance of the program. Furthermore, even if overrides are not cleared, information about past overrides is captured in multiple visual security indicators in the browser UI; these indicators assist users in taking informed decisions. One such indicator is a UI element called the *Site Identity Button* [24]. The *Site Identity Button* is comprised of a padlock icon at a bare-minimum. Figure 6 shows comparative screen shots of the site identity buttons in the Firefox OS browser and the Firefox desktop browser after a certificate override has taken place. As shown in the figure, the padlock icon is gray colored when a certificate override has taken place on the desktop browser. On clicking the padlock, an informative security dialog is displayed to the user which presents textual feedback on the state of connection security. While the browser app on Firefox OS has options which permit clearing browsing history, cookies, and stored data, none of these are tied to sanitizing temporary certificate overrides. An examination of the source code of Firefox OS' NSS port reveals that the option of overriding `Active Logins` is present. Evidently, this option is neither exposed in the browser app nor the certified Settings app.

The site identity button in Firefox OS' browser app has the following states:

---

[6]Removal of temporary certificate overrides is tied to clearing Active Logins from recent browsing history.

1) Web page has no security (HTTP), visually represented by the globe icon,
2) Page is insecure or displays mixed (HTTP and HTTPS) content, represented by a broken gray colored padlock,
3) Page is secure, represented by a green colored padlock.

In the Gecko port for Firefox OS, certificate overrides are incorrectly mapped to the secure state of the site identity button; the padlock is thus green colored (see Figure 6) even when a web page's SSL certificate has been manually overridden. Additionally, because of limited screen real estate, the security information dialog that is supposed to pop-up on clicking the site identity button is absent in the browser app. In the mozbrowser view that is commonly used by Firefox OS apps to render web pages, the site identity button is absent. Certificate validation or overrides in the mozbrowser view therefore go unnoticed. Given that temporary certificate overrides remain in Gecko's program memory for a substantial time-period, this poses a realistic threat. Tricking the user into overriding the certificate warning for a given web origin is sufficient to compromise subsequent visits to the affected web origin across multiple apps. This includes the default browser app, *mozbrowser* instances of apps belonging to the same domain, and third-party apps that load web content (e.g. for OAuth based authentication) from the affected web origin. This was manually verified using an experimental setup.[7]

*Threat Scenario:*

1) Victim connects to compromised network, such as public wifi.
2) Attacker performs MITM attack against HTTPS[8] and presents a fake certificate.
3) Victim overrides certificate warning temporarily, assuming the action impacts only the present page visit.
4) Page is loaded with green padlock in the location bar.
5) All subsequent connections can be MITM.

If device restart has not taken place, the MITM attack outlined in the preceding scenario might recur should the victim reconnect to the compromised wifi at a later time. On successive connections to a domain whose certificate has been overridden by the victim in the past, a full SSL handshake with the attacker's cached SSL certificate takes place silently i.e., the victim is neither presented with a warning interstitial, nor is a possible MITM signaled in the browser UI. Apart from initiating MITM attacks on newer connections, the adversary can steal cookies from the victim's unexpired sessions with the compromised domain across applications.

A malicious application could aid this process by detecting when a MITM attack is present. The application could provide the user with some plausible sounding reason for requiring a certificate override. Certificate warnings are already confusing, and any inconsistencies in sandboxing behavior only increase the cognitive attack surface.

### B. Application Code and Manifest Provisioning

While Mozilla ensures that packaged applications are digitally signed and fetched from the marketplace over SSL, hosted applications are provisioned as if they were normal web content. Because hosted apps still need to fetch application manifests and code from a web origin, Mozilla recommends that app developers serve them over HTTPS [25].

The application manifest file contains security sensitive information including a list of permissions requested by an app among other things. We collated an exhaustive list of hosted manifest URLs and noted that 92.8%[9] of hosted applications fetch manifest and application code over HTTP. We evaluated the threat of a man-in-the-middle intercepting a hosted manifest/application code and subsequently modifying it before relaying the modified content to the end-user. As proof of concept, we inserted the *audio-capture* (microphone) permission in the manifest of an application that is designed to only *play* audio content. Subsequently, while application code was being fetched from the remote server (over HTTP), we altered it by adding a record audio functionality to the app.

The threat of an active attacker modifying the manifest and application code in transit is constrained in two ways. Firstly, sensitive permissions available to hosted apps such as *audio-capture* and *geolocation* are explicit i.e., prompt on first use. Secondly, Mozilla segregates permission sets for hosted and privileged applications. Should an attacker request, say, the *Contacts* permission (a privileged permission on Firefox OS), Gaia's application installer component will trigger a signature check. Upon finding that the file in question is a manifest and not a signed zip package, the installation process is aborted. Unprivileged implicit permissions, however, can be injected without the user's knowledge. Since the *App Permissions* tab in the *Settings* app does not list implicit permissions, there is no way for a user to tell that the permission requests have been modified. Presently, implicit permissions for hosted apps include *audio*, *fmradio*, *alarms*, *desktop-notification*, and *storage* [26]. These are relatively benign permissions, but their absence from the permissions UI combined with the lack of an integrity mechanism for hosted application code would be a cause for concern should Mozilla enlarge the set of implicit permissions for hosted applications.

Moving forward, requiring that security sensitive hosted applications serve their manifest (and application code) over HTTPS would be a step in the right direction.

### VI. RELATED WORK

To the best of our knowledge, this paper is the first to look at the security of Firefox OS. The differences between Android and Firefox OS permission enforcement are discussed in [20], but the focus of that work is on Android.

While real-world studies [14], [15] have observed problems in SSL validation across popular web and smartphone applications and middleware libraries, they do not look at how certificate overrides are handled at the client side. On Firefox OS, SSL validation is done by system software; applications are not trusted to validate certificates on their own.

---

[7]We used mitmproxy (http://mitmproxy.org/) to intercept/modify HTTP(S) traffic from Geekphone's Peak phone running Firefox OS 1.4.0.0-prerelease.

[8]Web domains for which HTTP Strict-Transport-Security (HSTS) is hard-coded in the browser are not vulnerable to the described attack primarily because certificate overrides for these domains are disallowed.

[9]1106 out of a total of 1191 hosted applications returned by the marketplace API, as of 22nd February, 2014

Amrutkar et al. [10] perform an empirical evaluation of security indicators across mobile browser apps based on W3C guidelines for web user interface. They conclude that mobile web browsers implement only a subset of desktop browsers' security indicators, leaving mobile users vulnerable to attacks that the indicators are designed to signal. While our study does not intend to evaluate the state of security indicators in Firefox OS, our observations could be used as a starting point for extrapolating the findings in [10] to Firefox OS. Akhawe et al. [8] quantify the effectiveness of desktop browser warnings by measuring their click-through rates. Their finding that a high proportion of users click through SSL warnings, although from a desktop browser user base, lends credence to the attack scenario based on a certificate override that we demonstrate.

There are a number of JavaScript static analysis tools that were evaluated for their ability to analyze Firefox OS applications. Many could be adapted to serve the same purpose as ours, of which the three most practical are discussed here. WALA [6] provides a rich set of sophisticated analyses [33], but the JavaScript front-end had trouble parsing applications in the market, and it has a number of scalability issues. The results from WALA are clearly superior, but running it against thousands of applications, the majority of which include libraries such as jQuery, is computationally prohibitive. Our lightweight approach, in contrast, can analyze the entire market in a matter of minutes. JSPrime [27] is also capable of performing dataflow analysis, but the architecture is extremely difficult to extend. ScanJS [1], written in-house by Mozilla, is closest in spirit to our own. It is based on the same technologies as our tool, but looks only at AST patterns rather than performing dataflow analysis. It would not be difficult to extend ScanJS to perform our checks, and we are working with Mozilla to do so where it makes sense.

Dynamic taint analysis is a promising direction for future work, and has proven very effective for Android. The same approach taken by TaintDroid [13] could work for Firefox OS. Indeed, a recent project has introduced string-based taint support into the SpiderMonkey JavaScript engine [5]. It would be difficult to have code reviewers or developers use a tainted JavaScript engine, however, as it would require merging taint support into mainline Firefox or running a separate build of Firefox to use for Firefox OS development/review.

HTML injection is necessary for a number of the attacks discussed in this paper, which puts renewed emphasis on addressing the issue of structural integrity. Proposals such as Noncespaces [16] and Blueprint [21] address the issue, but fall short of being usable for Firefox OS. Noncespaces presupposes the use of XHTML, and both are aimed primarily at server-side web applications.

Carlini et al. [12] provide an evaluation of the Chrome extension security architecture. Through manual examination of a sample set, they observe that extensions are vulnerable to web attackers via HTML injection, and to network attackers because of the use of plaintext HTTP for XHR and fetching JavaScript. Their vulnerability results are similar to ours. Although Chrome extensions and Firefox OS apps share several characteristics—distribution through a centralized market, CSP restrictions, written in HTML/JavaScript—there are noteworthy differences. Chrome extensions make use of privilege separation between the portion of the extension that interacts with the DOM and the privileged core that makes browser API calls. Chrome also provides an "isolated world" which prevents web content from interacting directly with the extension. Firefox OS does not use privilege separation within an app, nor does it provide an isolated world. Extensions have a large attack surface resulting from their interactions with many pages, which partially motivates the introduction of these isolation mechanisms. Most Firefox OS apps, in contrast, interact with a limited number of other pages. Another difference is that the CSP for Chrome extensions can be relaxed. Firefox OS is less forgiving: privileged apps can only modify the CSP to become more restrictive. The CSP for Chrome packaged apps (not discussed in [12]), is actually more restrictive than Firefox OS, as it disallows external resources [2]. This stricter CSP would have prevented some of the vulnerabilities we found from being exploitable, as the exploits relied on the injection of an `iframe` with remote content. Chrome apps also differ from Firefox OS apps by the set of available APIs and permissions. Systematically classifying the similarities and differences of vulnerabilities between packaged web apps on a variety of platforms is an important effort that we leave to future work.

## VII. CONCLUSIONS

Firefox OS is the most serious attempt yet made by any modern operating system to merge the web and mobile experiences. This has led to a number of interesting security challenges as the web-based security model, firmly rooted in the same-origin policy, collides with the application-centric sandboxing associated with modern mobile operating systems. The web brings to mobile a number of security lessons learned the hard way, such as the importance of provenance in all communication channels. But it also brings risks, not the least of which being XSS attacks against highly privileged applications. Mozilla takes several proactive measures to address these risks: application code review, a strong Content Security Policy (CSP) for privileged applications, and application sandboxing to name a few. We find that these measures have improved upon both the web and mobile security models, but that there is still room to grow.

Firefox OS is in a transition phase, as it still relies on libraries that are firmly entrenched in the desktop security model. Services that used to be tightly coupled to the browser can wind up conflicting with the use of individual applications as security principals. Moreover, the outmoded desktop notion of application lifecycle can come into conflict with an always-on browser. We show one example of how a known Firefox OS issue can be traced back to these conflicts. This is a class of bugs that has not yet been well-explored, and we suspect that more issues will arise along these lines.

The Firefox OS market is small, but steadily growing. As the market scales up, increased pressure will be put on reviewers of privileged applications. Even now, there exist quite a few applications that violate the security guidelines put forth by Mozilla. Some of these guidelines can be very easily checked. Flagging applications that fail to validate the origin of postMessage messages takes very little analysis, yet we find that the majority of applications that use postMessage fail to even look at the origin of messages. This leads us to the conclusion that even a small increase in the level of sophistication used to validate applications that come into the market would show appreciable results.

While app developers cannot be relied upon, and should not necessarily be forced to, correctly implement security checks, the structure of application code can be used to infer certain semantic features about the application. The presence of filtering a certain type of data might imply that the developer feels that type data might be controlled by an attacker, which is an otherwise undecidable problem. We have shown that in at least one instance, this led to the discovery of an otherwise unknown vulnerability. Though this is certainly a small result set, there appears to be promise that this approach could help prevent other vulnerable applications from entering the market.

ACKNOWLEDGEMENTS

This work was partially supported by the EU FP7 Trustworthy ICT program (FP7-ICT-2011.1.4) under grant agreement no. 317888 (project NEMESYS).